\documentclass{aa}

\usepackage{txfonts}
\usepackage{graphicx}

\begin{document}

\title{Centimetre continuum emission from young stellar objects in 
Cederblad~110}

\author{ K.\ Lehtinen\inst{1} \and J.\ Harju\inst{1} \and 
         S.\ Kontinen\inst{1} \and James L.\ Higdon\inst{2}}

\institute{Observatory, T\"ahtitorninm\"aki, P.O.\
           Box 14, 00014, University of Helsinki, Finland
           \and
           Department of Astronomy, 215 Space Sciences Building, 
           Cornell University, Ithaca, NY 14853-6801, 
           USA}

\date{Received / accepted}

\offprints{K.\ Lehtinen (kimmo.lehtinen@helsinki.fi)}

\titlerunning{Continuum observations of Cederblad~110}
\authorrunning{K.\ Lehtinen \& J.\ Harju \& S.\ Kontinen \& J.L.\ Higdon} 

\abstract{The low-mass star formation region associated with the
reflection nebula Cederblad~110 in the Chamaeleon~I cloud was mapped
with the Australian Telescope Compact Array (ATCA) at 6 and 3.5\,cm.
Altogether 11 sources were detected, three of which are previously
known low mass young stellar objects associated with the nebula: the
illuminating star IRS~2 (Class~III, {\it Einstein} X-ray source
CHX~7), the brightest far-infrared source IRS~4 (Class~I), and the
weak X-ray source CHX10a (Class~III).  The other young stellar objects
in the region, including the Class~0 protostar candidate Cha-MMS1,
were not detected.  The radio spectral index of IRS~4 ($\alpha = 1.7
\pm 0.3$) is consistent with optically thick free-free emission
arising from a dense ionized region, probably a jet-induced shock
occurring in the circumstellar material.  As the only Class~I
protostar with a 'thermal jet' IRS~4 is the strongest candidate for
the central source of the molecular outflow found previously in the
region. The emission from IRS~2 has a flat spectrum ($\alpha = 0.05
\pm 0.05$) but shows no sign of polarization, and therefore its origin
is likely to be optically thin free-free emission either from
ionized wind or a collimated jet. The strongest source detected in
this survey is a new compact object with a steep negative spectral
index ($-1.1$) and a weak linear polarization ($\sim 2 \%$), which
probably represents a background radio galaxy.

}

\maketitle

\section{Introduction}

The Cederblad~110 (Ced~110) reflection nebula in the Chamaeleon~I
molecular cloud is associated with one of the nearest low-mass star
forming regions, with young stellar objects (YSOs) in all different
phases from a pre-stellar clump and Class~0 protostellar candidate to
a Class~III pre-main sequence star (for YSO classification, see Lada
(\cite{lada87}) and Andr\'e et~al.\ (\cite{andre93})).  The region is
therefore favourable for studies of the dependence of radio continuum
emission on the stellar evolutionary stage. On the other hand, a
centimetre continuum mapping may reveal protostars, which have
remained undetected in the infrared because of large optical depths.
With these prospects as motivation we have mapped the central part of
the nebula with the Australian Compact Array (ATCA) in continuum at
6\,cm and 3.5\,cm, in order to investigate the nature of the known
stellar sources and to search for new protostellar candidates.

The region has been studied in great detail in the near-IR by
Zinnecker et~al.\ (\cite{zinnecker99}) and Persi et~al.\
(\cite{persi01}), in the mid-IR by Persi et~al.\ (\cite{persi00}), and
in the far-IR by Lehtinen et~al.\ (\cite{lehtinen01}). These
papers contain all the relevant references to previous studies at
various wavelengths ranging from 1.3\,mm to X-rays. The YSOs
identified in different surveys are listed in Table~\ref{table:ysos},
which also gives the estimated infrared class, and the region of the
electromagnetic spectrum where the source is detected.  The brighest
star in visible light and in the X-rays is the weak-lined T~Tauri star
(Class~III) IRS~2 ({\sl Eintein} X-ray source CHX~7), which is the
illuminating star of Ced~110. The Class~I object IRS~4 is the most
prominent far-infrared source, whereas the 1.3\,mm continuum map is
dominated by the Class~0 candidate Cha-MMS1 (the references are given
in Table~\ref{table:ysos}).

The distribution of molecular gas in the region has been studied by
Mattila et~al.\ (\cite{mattila89}), and more recently by Mizuno
et~al.\ (1999).  Mattila et~al.\ (\cite{mattila89}) detected a bipolar
molecular outflow originating somewhere in the centre of Ced~110.
Reipurth et~al.\ (\cite{reipurth96}) mapped the centre of Ced~110
in continuum at 1.3\,mm with a resolution of 22\arcsec, and detected a
marginally resolved bright dust continuum source they called
Cha-MMS1. This source is coincident with the centre of a dense
molecular clump studied in several molecular species by Kontinen
et~al.\ (\cite{kontinen00}).  Reipurth et~al.\ suggested that
Cha-MMS1 contains a Class~0 protostar, which is the driving source of
the molecular outflow and Herbig-Haro objects HH~49/50 located about
10\arcmin south of it. Since the central sources of Herbig-Haro flows
are often detected in radio continuum (e.g.\ Rodr\'{\i}guez \&
Reipurth \cite{rodriguez96}), one of the aims of the observations
discussed here was to verify - or invalidate - this suggestion.

In Sect.~2 of this paper we discuss briefly the mechanisms of radio
emission associated with newly born stars.  In Sect.~3 we describe our
observational procedure. In Sect.~4 we present the results and discuss
the properties of individual sources in the light of previous surveys
and some theoretical expectations. Finally, in Sect.~5 we summarize
our conclusions.

\section{Cm-wavelength continuum emission from YSOs}

A large fraction of the youngest protostars, i.e.\ those belonging to
the Classes 0 and I, have been detected in radio continuum at cm
wavelengths (e.g.\ Anglada \cite{anglada96};
Rodr\'{\i}guez \cite{rodriguez94}). Class~III objects are often
associated with non-thermal radio emission.  On the other hand, the
detection rate amongst Class~II protostars is lower than for the other
classes.  Gibb (\cite{gibb99}) discussed this problem and suggested a
scenario in which the dominant emission mechanism at cm wavelengths
changes with the age of the YSO.  According to this Class~0 and
Class~I objects, which are in the main and late accretion phase,
respectively, emit thermal free-free emission from ionized jets, but
this emission declines with time when the accretion rate or outflow
efficiency (which are intimately linked) go down.  Finally in the
Class~III stage, synchrotron emission from the exposed
pre-main-sequence star becomes the dominant mechanism (see also
Wilking et~al.\ \cite{wilking01}).

Thermal emission can arise directly from stellar winds or collimated
jets, or from shocks associated with jets or infall onto accretion
disks. Non-thermal emission is related to magnetic fields close to the
chromospherically active YSO, and/or to the star-disc interaction
region. It is known that ionization by stellar photons is insufficient
to produce the continuum emission in low luminosity sources such as
those in Cederblad~110 (see e.g.\ Rodr\'{\i}guez et~al.\
\cite{rodriguez89}). Reviews on possible thermal and non-thermal
emission mechanisms in connection with young stars at different stages
of evolution are given e.g. in Anglada (\cite{anglada96}), Andr\'{e}
(\cite{andre96}), Panagia (\cite{panagia91}), and Feigelson \&
Montmerle (\cite{feigelson99}).  Models explaining the observed
properties of the free-free emission sources are presented e.g.\ in
Reynolds (\cite{reynolds86}, collimated ionized winds), Curiel et~al.\
(\cite{curiel87}, Herbig-Haro shocks), Neufeld \& Hollenbach
(\cite{neufeld96}, accretion shocks), and Ghavamian \& Hartigan
(\cite{ghavamian98}, shocks in dense gas).

\begin{table}
\caption[]{YSOs associated with Ced~110}
\begin{tabular}{lllll}
IR Class   & designation  & other name   & spectrum & references   \\ \hline
           &              &              &            &               \\
Class 0    &              &              &            &               \\
or earlier & Cha-MMS1     &              & 1.3\,mm    &    c,f        \\
           &              &              &            &               \\
Class I    &              &              &            &               \\
           & IRS~4        & ISO-ChaI~84  & IR         &    b,e,f,g,h  \\ 
           & IRS~6        & ISO-ChaI~92  & IR, X-ray  &    b,d,e,f,g  \\
           & ISO-ChaI~86  & NIR~98       & IR         &    e,g,f      \\
           & NIR~89       &              & NIR        &    g          \\
           &              &              &            &               \\
Class II   &              &              &            &               \\
           & ISO-ChaI~97  &  NIR~131     & IR         &    e          \\
           & ISO-ChaI~101$^1$
                          &  NIR~141     & IR, X-ray  &    d,e        \\
           & NIR~72       &              & NIR        &    g          \\
           & NIR~84       &              & NIR        &    g          \\
           &              &              &            &               \\
Class III  &              &              &            &               \\
           & IRS~2        & CHX~7,       & IR, X-ray  &    a,b,d,e,f  \\
           &              & ISO-ChaI~75  &            &               \\
           & CHX10a       & ISO-ChaI~117 & IR, X-ray  & a             \\
           &              &              &            &        \\ \hline
\end{tabular}
a Feigelson \& Kriss (\cite{feigelson89});
b Prusti et~al.\     (\cite{prusti91});
c Reipurth et~al.\   (\cite{reipurth96});
d Carkner et~al.\    (\cite{carkner98});
e Persi et~al.\      (\cite{persi00});
f Lehtinen et~al.\   (\cite{lehtinen01});
g Persi et~al.\      (\cite{persi01});
h Zinnecker et~al.\  (\cite{zinnecker99}) \\
$^1$ The infrared spectral index between 2.2\,$\mu$m and 14.3\,$\mu$m, 
defined as
$\alpha_{\mathrm IR} \equiv \mathrm{d} \log(\lambda F_{\lambda}) / 
\mathrm{d} \log(\lambda)$, is -0.2. This value is close to the border 
between Class~I and II objects (-0.3)
\label{table:ysos}
\end{table}

\section{Observations and data reduction}

Cederblad~110 was observed with the Australia Telescope Compact Array
(ATCA), located at Narrabri, New South Wales, Australia. The
observations were made on 2000 May 24 and 28, with the 1.5A array
configuration. The frequencies 4.80\,GHz (6\,cm) and 8.64\,GHz
(3.5\,cm) were observed simultaneously. The correlator configuration
used provided at each frequency a total bandwidth of 104\,MHz recorded
as a 13 channel spectrum.

The calibration sources were selected from the ATCA Calibrator Source
Catalogue (Reynolds \cite{reynolds97}).  The Seyfert~2 galaxy
PKS~1934--638 was used as the primary flux calibrator, and the
phase-reference calibrator was the quasar PKS~1057--797.  The flux
densities and the degrees of polarization of 1057--797 were found to be
higher than those given in ATCA catalogue (see
Table~\ref{table:observations}).

\begin{table}
\caption[]{Observational parameters. The position angle is measured to
East from North}
\begin{tabular}{lll}
Target source       & Ced~110 &                                            \\
Phase centre        & R.A.\ $11^{\rm h}06^{\rm m}39.1^{\rm s}$ 
                    & Dec.\ $-77^\circ 22^{\prime} 58^{\prime\prime}$      \\
                    & (J2000.0)  & (J2000.0)                               \\
Frequency           &   4.80\,GHz  & 8.64\,GHz                             \\ 
Bandwidth           & 104 ($13 \times 8$) MHz & 104 ($13 \times 8$) MHz    \\ 
                    & &                                                    \\
Flux calibrator     & PKS~1934--638 &                                      \\ 
                    & $S_{4.80 \rm GHz} = 5.83$\,Jy
                    & $S_{8.64 \rm GHz} = 2.84$\,Jy                        \\
                    &     &                                                \\ 
Phase calibrator    & PKS~1057--797 &                                      \\
                    & $S_{4.80 \rm GHz} = 2.58$\,Jy
                    & $S_{8.64 \rm GHz} =  2.44$\,Jy                       \\
                    & Linear polarization &                                \\ 
                    & 6\% (P.A.\ $80^\circ$) & 7\% (P.A.\ $83^\circ$)      \\
Image processing    & Naturally weighted  &                                \\
Synthesized beam    & $6.4\arcsec\times 5.8\arcsec$  
                    & $4.0\arcsec\times 3.7\arcsec$                        \\ 
Beam position angle &  $-48^\circ$
                    &  $-58^\circ$                                 \\
RMS noise level     & $34\,\mu$Jy\,beam$^{-1}$  & $35\,\mu$Jy\,beam$^{-1}$ \\  
\end{tabular}
\label{table:observations}
\end{table}

The observations were made in two 12 hours runs, alternating between
the target source (25 minutes) and the phase calibrator (5 minutes).
During the first night the observations were interrupted by rain for
an hour, but during the second night the weather was excellent.  

The data were calibrated and cleaned with the Miriad package (Sault
et~al.\ \cite{sault95}). The inversion was performed using all the
channel information (i.e.\ multifrequency synthesis).  The flux
densities of the detectected sources were derived from naturally
weighted images corrected for the primary beam responses.  The noise
levels at the centres of the images, and the synthesized beams are
given in Table~\ref{table:observations}.  Intensity maxima exceeding
$3\times$ the local rms are considered as detections.

\section{Results and discussion}

Altogether 11 radio sources were identified by inspecting the maps
visually.  The characteristics of the detected sources are summarized
in Table~\ref{sources}. In this table we give for each source the
designation, the equatorial coordinates, the angular size, the flux
densities at 3.5 and 6\,cm and the spectral index, $\alpha$,
determined from the latter two values.  The radio source Ced~110~R5
has two maxima, which are treated as separate components. The error in
the $\alpha$ is estimated on the basis of random noise only.  Objects
that may be associated with the radio sources at other wavelengths
were searched in the SIMBAD database.

\begin{table*}
\caption[]{Radio continuum sources detected in Cederblad~110.  Flux
upper-limits are calculated assuming a point source with a peak
intensity of three times the rms.  The positional accuracy is expected
to be better than 2$\arcsec$.  The angular size is the
measured FWHM along the major and minor axes at 3.5\,cm, or at 6\,cm if
3.5\,cm data are not available, with uncertainty in parentheses.  The
quantity $S$ is the intensity integrated over the source. The errors
in $S$ and $\alpha$ are estimated on the basis of the local rms noise.}
\begin{flushleft}
\begin{center}
\begin{tabular}{rrrrrrrr}
\hline\noalign{\smallskip}
  &  \multicolumn{2}{c}{Position}  &  Angular  &  $S$(3.5\,cm)  &  
     $S$(6\,cm)  & $\alpha$  & Notes	 			          \\
Source   &  $\alpha$(2000)  &  $\delta$(2000)  &  size [$\arcsec$]  & 
[mJy]  &  [mJy]  &  &	 						  \\
\noalign{\smallskip}
\hline\noalign{\smallskip}
Ced~110~R1 & $11^{\rm h} 05^{\rm m} 13\fs9$
 & $-77\degr 21\arcmin 01\farcs1$ & 4.0$\times$4.1 (0.1, 0.1)
 & 7.8$\pm$0.1 & 14.74$\pm$0.02 & -1.09$\pm$0.03  & 1 \\
Ced~110~R2 &  $11^{\rm h} 05^{\rm m} 44\fs5$
 & $-77\degr 26\arcmin 39\farcs3$ & 5.9$\times$10.7 (0.1, 0.1)
 & $<1.3$ & 1.6$\pm$0.1 & $<-0.3$  & 2 \\
Ced~110~R3 &  $11^{\rm h} 06^{\rm m} 00\fs8$
 & $-77\degr 22\arcmin 05\farcs9$ & 3.1$\times$4.2 (0.3, 0.4)
 & 0.12$\pm$0.04 & $<0.06$ & $>1.3$  & 2 \\
IRS~2 & $11^{\rm h} 06^{\rm m} 15\fs4$
 & $-77\degr 21\arcmin 56\farcs8$ & 3.6$\times$3.9 (0.1, 0.1)
 & 1.10$\pm$0.02 & 1.08$\pm$0.01 & 0.05$\pm$0.05  & 1 \\
Ced~110~R4 &  $11^{\rm h} 06^{\rm m} 17\fs4$
 & $-77\degr 22\arcmin 38\farcs5$ & 3.2$\times$4.2 (0.6, 0.8)
 & 0.04$\pm$0.02 & $<0.03$ & $>0.1$  & 2 \\
IRS~4 & $11^{\rm h} 06^{\rm m} 46\fs5$
 & $-77\degr 22\arcmin 33\farcs2$ & 3.5$\times$4.6 (0.3, 0.4)
 & 0.17$\pm$0.02 & 0.06$\pm$0.01 & 1.7$\pm$0.3  & 1 \\
Ced~110~R5a & $11^{\rm h} 07^{\rm m} 13\fs4$
 & $-77\degr 21\arcmin 57\farcs9$ & 5.4 (0.3)
 & $<0.12$ & 0.21$\pm$0.02 & $<-0.9$  & 1 \\
Ced~110~R5b & $11^{\rm h} 07^{\rm m} 12\fs2$
 & $-77\degr 21\arcmin 55\farcs7$ & 5.3 (0.4)
 & $<0.12$ & 0.15$\pm$0.02 & $<-0.4$  & 1 \\
Ced~110~R6 & $11^{\rm h} 07^{\rm m} 49\fs5$
 & $-77\degr 21\arcmin 54\farcs6$ & 5.7$\times$6.9 (0.3, 0.4)
 & $<0.22$ & 0.15$\pm$0.04 & $<0.7$  & 2 \\
CHX10a &  $11^{\rm h} 07^{\rm m} 55\fs7$
 & $-77\degr 27\arcmin 25\farcs0$ & 6.5$\times$13.7 (0.3, 0.6)
 & - & 0.45$\pm$0.11 & - & 2   \\
Ced~110~R7 &  $11^{\rm h} 08^{\rm m} 08\fs0$
 & $-77\degr 23\arcmin 47\farcs1$ & 6.0$\times$8.9 (0.1, 0.1)
 & $<1.3$ & 1.33$\pm$0.09 & $<0.0$  & 2 \\
\noalign{\smallskip}
\hline
\end{tabular}
\end{center}
\end{flushleft}
$^1$ Flux is the peak flux of a gaussian fit                         \\
$^2$ Flux is derived by summing up the pixels having flux greater 
     than the local 3\,$\sigma$ noise
\label{sources}
\end{table*}         

\begin{figure}
\resizebox{\hsize}{!}{\includegraphics{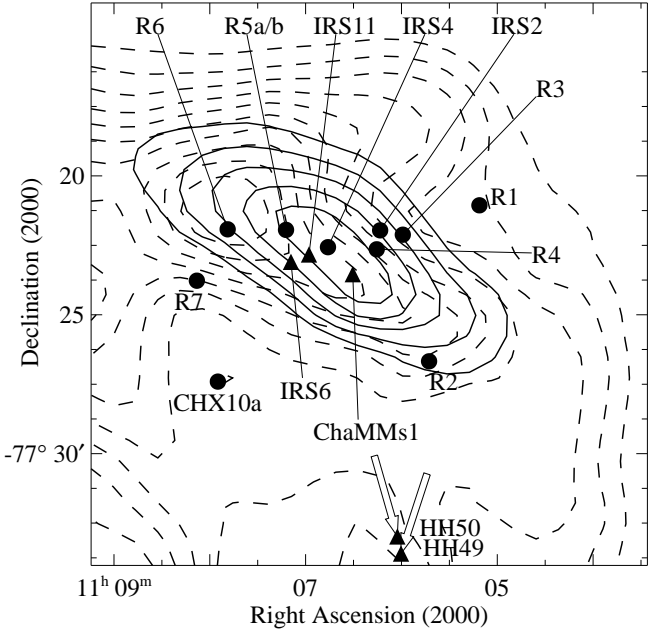}}
\caption{Radio and infrared sources in the Cederblad~110.  The cm
continuum sources are marked with filled circles and the sources
detected in other surveys are marked with filled triangles. Solid
contours show the intensity of the 200\,$\mu$m emission ridge
(Lehtinen et~al.\ \cite{lehtinen01}), from 40 to 115 in steps of
15\,MJy\,sr$^{-1}$.  The dashed contours show the integrated intensity of
C$^{18}$O($J$=1--0), from 0.2 to 2.0 in steps of 0.2\,K\,km\,s$^{-1}$.
The velocity vectors of Herbig-Haro~49 and 50 (mean of HH~50a--e) are
indicated with arrows}
\label{fig1}
\end{figure}

\begin{figure}
\resizebox{\hsize}{!}{\includegraphics{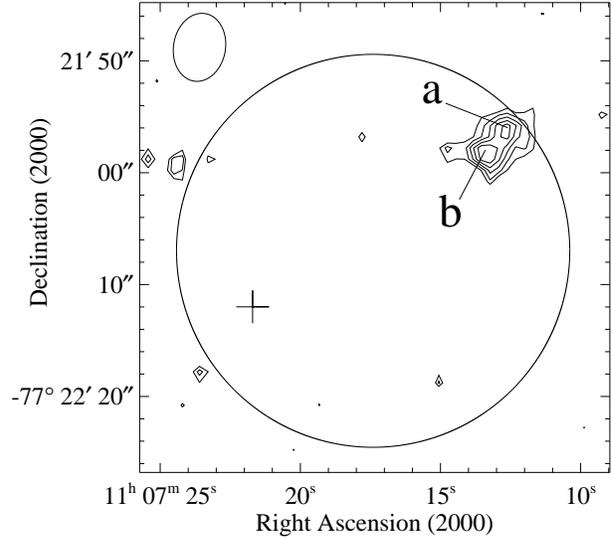}}
\caption{The 6\,cm contour map of the source Ced~110~R5. The contours
are from 70\,$\mu$Jy\,beam$^{-1}$ to 150\,$\mu$Jy\,beam$^{-1}$ in
steps of 20\,$\mu$Jy\,beam$^{-1}$. The position of the infrared source
ISO-ChaI~101 (the cross) and a circle depicting the X-ray intensity
contour of 25 counts per pixel measured with ROSAT PSPC (Carkner
et~al.\ \cite{carkner98}) are shown. The FWHM beam size is shown at
top left}
\label{fig2}
\end{figure}

Three of the detected sources are previously known objects: the
infrared sources IRS~2 and IRS~4, and the X-ray source CHX10a. The
rest are new sources, and we designate them as Ced~110~R1, Ced~110~R2,
etc., where ``Ced~110~R'' stands for ``Ced~110 radio source''.  The
mm-wave continuum source Cha-MMS1 was not detected.  As can be seen
from Table~\ref{sources}, only three sources, Ced~110~R1, IRS~2 and
IRS~4, are visible at both 3.5 and 6\,cm.

The positions and sizes of the detected sources have been determined
by two-dimensional gaussian fits. If a source is pointlike, the flux
and its error given in Table~\ref{sources} are the peak flux and its
error of the gaussian fit.  For resolved sources the flux is
determined by adding up the pixels with flux density greater than the
local 3\,$\sigma$ rms noise value.  The error of the integrated flux
density for resolved sources is derived using the formula
\begin{equation}
\sigma = \sigma_0 (N/\theta)^{1/2} \; ,
\end{equation}
where $\sigma_0$ is the rms noise in the background, $N$ is the number
of pixels over which the source is summed up, and $\theta$ is the
number of pixels per beam.  

If a source is detected at one frequency only and is found to be
extended, the upper limit of integrated flux at the other frequency is
estimated according to the formulation of Bertl\'an et~al.\
(\cite{beltran01}).  We use the equation $S_{\mathrm{lim}}=3 \sigma
A^{0.7}$, where $A$ is the source size in units of beam area, $\sigma$
is the local rms noise, and assume a nominal $3\sigma$ detection as
for point sources. 

The locations of the radio sources are indicated with filled circles
on the map presented in Fig.~\ref{fig1}. This figure shows the
extended far-IR emission ridge from Lehtinen et~al.\
(\cite{lehtinen01}) overlaid on an unpublished C$^{18}$O$(J=1-0)$
integrated intensity map. The sources detected previously in other
surveys are indicated with filled triangles.

In the following we first discuss spectral indices predicted by
different models of thermal radio emission from YSOs, and then
in the next subsection, the nature of individual sources
detected in this survey on the basis of their radio properties and
association with known objects.

\subsection{Spectral indices of thermal sources}

The predicted spectral index $\alpha$ ($S_\nu \propto \nu^\alpha)$ of
the so called standard spherical wind, i.e.\ a fully ionized, constant
velocity and isotropic stellar wind, is 0.6 at centimetre wavelengths
(e.g.\ Panagia \& Felli \cite{panagia75}).  Reynolds
(\cite{reynolds86}) have modelled continuum emission from both
collimated and accelerated spherical flows. According to these results
the spectral index can increase above 0.6 for an accelerated flow, but
a collimated ionized flow produces a flat spectrum with $\alpha <
0.6$.  Rodr\'{\i}guez et~al.\ (\cite{rodriguez93}) have shown that the
spectral index expected from an ionized jet is always greater than
$-0.1$, irrespective the geometry and possible clumpiness.
 
Optically thick free-free emission can be detected from shocks
occurring in very dense gas, as in the situations of accreting material
falling onto a circumstellar disk or a jet penetreting into dense
circumstellar material (Neufeld \& Hollenbach \cite{neufeld96};
Ghavamian \& Hartigan \cite{ghavamian98}).

\subsection{Radio emission from the young stars IRS~2, IRS~4 and CHX10a, 
and thermal sources possibly associated with Ced~110}

{\sl IRS~2} is the second strongest 3.5\,cm source in this survey, and
the third strongest at 6\,cm. The spectral index is $0.05 \pm 0.05$,
and the radio continuum luminosity at 3.5\,cm $L_{3.5 \rm cm} =
0.025\,{\rm mJy\,kpc^{2}}$.  The luminosities are estimated using the
distance 150 pc to Chamaeleon I (Knude \& H{\o}g \cite{knude98}).  As
an evolved YSO IRS~2 is probably not associated with a thick accretion
disk.  The ionized winds or collimated jets emanating from this star
are therefore likely to proceed freely, and according to models quoted
in Sect.~2 the free-free emission should show a flat spectrum as
indeed is the case. Consider the model of Reynolds (\cite{reynolds86})
for an isothermal, constant-velocity, fully ionized flow with $w(r)
\propto r^{\epsilon}$, where $w(r)$ is the half-width of the jet
perpendicular to its axis and $r$ is the distance from the central
source. In this model the spectral index in the optically thick case
depends on $\epsilon$ according to $\alpha = 1.3-0.7/\epsilon$. The
spectral index observed towards IRS~2 would then correspond to the
collimation exponent $\epsilon = 0.6$.

The powerful X-ray emission associated with IRS~2 implies, on the
other hand, magnetic activity close to its surface. Magnetic loops or
fast shocks caused by ejecta from the surface may be sources of
non-thermal radio emission. The observed spectral index is in fact
consistent with moderately optically thick synchrotron emission. These
kind of activities are generally associated with detectable circular
polarization.  No polarized signal is, however, detected towards this
source. Without further evidence for the synchrotron model we think
the more likely alternative is that the radio spectrum is dominated by
free-free emission from ionized jets.

\vspace{3mm}

{\it IRS~4}: The radio source is located about 2\arcsec to the
northwest from the position of the highly reddened illuminating star
of the bipolar reflection nebula discovered by Zinnecker et~al.\
(\cite{zinnecker99}) (see also Persi et~al.\ \cite{persi01}). Taking
the positional inaccuracies into account, the radio source is almost
certainly associated with this star.  The spectral index, $1.7\pm0.3$,
indicates optically thick free-free emission, which according to the
models of Neufeld \& Hollenbach (\cite{neufeld96}) and Ghavamian \&
Hartigan (\cite{ghavamian98}) signifies shocks in dense circumstellar
material.  The existence of a massive circumstellar disk or envelope
around IRS~4 is evident from the near-IR images of Zinnecker et~al.\
(\cite{zinnecker99}) and Persi et~al.\ (\cite{persi01}). Far-IR
ISOPHOT observations of Lehtinen et~al.\ (\cite{lehtinen01}), where
IRS~4 is by far the strongest source in the region, indicate a large
circumstellar mass. According to the estimates of Persi et~al.\ the
mass of the central object is very low ($0.1-0.3\,M_\odot$), which
suggests that powerful accretion shocks are less likely.  Interaction
between stellar jets and surrounding material are therefore the most
plausible origin of the radio continuum emission for this object.

IRS~4 lies close to the centre of the molecular outflow detected in
$^{12}$CO by Mattila et~al.\ (\cite{mattila89}).  As the only Class~I
object in this region provedly associated with thermal jets IRS~4 is
probably the central source of the outflow.

\vspace{3mm}

{\it CHX10a}: Weak emission at 6-cm originates close to the X-ray
source CHX10a, which is classified as a weak-line T~Tauri star, i.e.\
a Class~III object (Feigelson \& Kriss \cite{feigelson89}; Feigelson
et~al.\ \cite{feigelson93}). In the latter study the source was
designated as CHXR28 (CHXR is for ``Chamaeleon I X-ray ROSAT''), and
it is also known as the infrared source ISO-ChaI~117 (Persi et~al.\
\cite{persi00}). At 3.5\,cm the source lies outside the field of view
limited by the primary beam response, and unfortunately we cannot
derive even the upper limit for the spectral index.  
Judging from the nature of star, the 6\,cm emission is likely to
be synchrotron emission originating from magnetic loops close to the
stellar surface.

\vspace{3mm}

{\it Ced~110~R3 and 4}: These weak sources lying close to IRS~2 in the
centre of Ced~110 are detected at 3.5\,cm only.  The lower limits of
their spectral indices are consistent with thermal emission. We
therefore consider it possible that these sources are associated with
the Chamaeleon I cloud, perhaps representing shocks associated with
jets emanating from one of the young stars in the region.

\subsection{Non-detections}

{\it Cha--MMS1}: A Class~0 object should, by definition, show indirect
evidence for the presence of a central protostellar object in the form
of cm-wavelength continuum emission or a molecular outflow.  Since
Cha-MMS1 was not detected in either frequency band, and the centre of
the molecular outflow is not coincident with it, Cha-MMS1 does not
seem to fulfil the criteria for a Class~0 protostar. Probably it
represents an earlier stage where violent processes have not yet taken
effect. The empirical relation of Harvey et~al.\ (\cite{harvey02})
between the 3.6\,cm continuum flux $S_{\mathrm{3.6\,cm}}$ and
bolometric luminosity $L_{\mathrm{bol}}$ for Class~0 and I
protostellar sources gives an upper limit of $L_{\mathrm{bol}} \la
0.17 \mathrm{L_{\sun}}$, when the 3\,$\sigma$ upper limit for the
3.5\,cm flux density is used.  The bolometric luminosity
$\sim0.45\,L_{\sun}$ derived from far-IR observations (Lehtinen
et~al.\ \cite{lehtinen01}) is higher than this upper limit, suggesting
that the Harvey et~al.\ relation is not valid for Cha-MMS1, which
supports the idea that it does not belong to Class~0.

\vspace{3mm}

{\it IRS~6}: The Class~I object IRS~6 belongs to the most luminous
YSOs in the region along with IRS~2 and IRS~4. Persi et~al.\
(\cite{persi01}) resolved this object into two components, IRS~6a and
6b, of which 6a dominates the mid-IR and far-IR emission.  The far-IR
and sub-mm data suggest that the negative result at cm wavelengths
reflects profound differences between IRS~6 and IRS~4, and not merely
temporal variability (e.g.\ Lucas et~al.\ \cite{lucas00}). The far-IR
spectrum of IRS~6 indicates much weaker circumstellar dust emission
than in the case of IRS~4 (see Table~2 of Lehtinen et~al.\
\cite{lehtinen01}). Moreover, as discussed in Carkner et~al.\
(\cite{carkner98}), the fact that (unlike IRS~4) it was not detected
in the 1.3\,mm survey of Henning et~al.\ (\cite{henning93}) betokens a
lack of circumstellar envelope, and perhaps a more advanced stage of
protostellar evolution. The non-detection in the radio could then be
understood in terms of the evolutionary scenario laid out by Gibb
(\cite{gibb99}), and summarized in Sect.~2.

\subsection{Background objects}

The deep optical and near-IR surveys performed towards Ced~110 (Persi
et~al.\ \cite{persi01}) have probably revealed all Class~III protostars
in the region, for example the ``naked YSO'' -type objects with
non-thermal radio spectra detected by Andr\'e et~al.\ (\cite{andre92})
in the $\rho$~Ophiuchi cloud.  Radio sources with negative spectral
indices in our maps without any counterpart in the near-IR are
therefore likely to be background objects, and as we are rather far
from the Galactic plane, these objects are probably radio galaxies,
although pulsars cannot be totally excluded.

\vspace{3mm}

{\it Ced~110~R1}: The previously unknown compact source Ced~110~R1 is
by far the strongest 6\,cm and 3.5\,cm source in this survey.
The rather large negative spectral index, $\alpha=-1.09 \pm
0.03$, can be found among the various classes of extragalactic
steep-spectrum sources, but would not be unusual for pulsars, which
often have, however, still steeper spectra in this wavelength range
(see e.g.\ Kaplan et~al.\ \cite{kaplan00}). Ced~110~R1 can be detected
also in the Stokes Q and U images (220 and 160\,$\mu$Jy, respectively),
but not in Stokes V. As the I flux density is about 14.7\,mJy, these
results indicate linear polarization at the level 1.8\%.  Further
observations to investigate the nature of this bright object seem
warranted.

\vspace{3mm}

{\it Ced~110~R5}: This marginally resolved source detected at 6\,cm is
located within the dust emission ridge (see Fig.~\ref{fig1}), close to
the grouping of YSOs containing IRS~2, 4, 6 and 11.  A contour image
of Ced~110~R5 is presented in Fig.~\ref{fig2}.  The image shows two
emission maxima (components a and b) separated by $\sim 3\arcsec$, and
the southern source has an extension at about 2--3$\sigma$ level
reaching out to the east.  The parameters of the components given in
Table~\ref{sources} were determined by fitting two gaussian functions
to the image.  Both sources have negative spectral indices.  The
object lies close to the X-ray source which Carkner et~al.\
(\cite{carkner98}) associated with the infrared source ISO-ChaI~101
(see their Fig.~1).  The angular distance to Ced~110~R5 from centre of
the X-ray source ($\sim 15\arcsec$) is about the same as from there to
ISO-ChaI~101 (see Fig.~\ref{fig2}). Judging from the positional
coincidence the association of Ced~110~R5 with the X-ray source is
equally likely as it is for ISO-ChaI~101. The separation between
between Ced~110~R5 and ISO-ChaI~101 (corresponding to 4000\,AU at a
distance of 150\,pc) is too large to expect a connection.  Therefore,
as it is not associated with an infrared source, Ced~110~R5 is
probably a background radio galaxy.

\vspace{3mm}

{\it Ced~110~R2, Ced~110~R6 and Ced~110~R7}: These sources are visible
at 6\,cm only and for R2 and R7 the upper limits of the spectral
indices are negative, suggesting optically thin synchrotron emission.
The large upper limit of the spectral index of R6 is due to its
location close to the outer edge of the map.  Most likely these
sources background radio galaxies. Ced~110~R2 and Ced~110~R7 are
slightly extended with elliptical shapes. Ced~110~R2 is the second
brightest source at 6\,cm.  It lies at a distance of about $10\arcsec$
from the X-ray source CHXR~15, which Feigelson et~al.\
(\cite{feigelson93}) identified with a star visible on the ESO/SERC
Sky Survey plates with an R-band magnitude of 16.0. The radio sources
Ced~110~R6 and Ced~110~R7 have no counterpart in other surveys. The
former object lies in the direction of the extended 200\,$\mu$m
emission ridge, but the location is probably purely coincidental.

\section{Conclusions}

Three YSOs associated with the Cederblad~110 star formation region
have been detected with the Australian Telescope Compact Array at
3.5\,cm and 6\,cm. These three sources, IRS~4 (Class~I), IRS~2
(Class~III) and CHX10a (Class~III) represent different stages of
stellar evolution.  IRS~2 shows a flat spectrum but no polarization,
and these properties can be explained by optically thin free-free
emission from ionized stellar wind or a collimated jet. The spectral
index of IRS~4 is consistent with optically thick free-free emission,
probably arising from a jet-driven shock in the circumstellar
material. The other luminous Class~I source IRS~6 is not detected in
radio, but there are indications in previous sub-mm and far-IR
observations that it has less circumstellar material than IRS~4 and
can thus represent a more evolved protostar.  The fact that of the
four Class~I objects in the region only the strongest far-IR source
(IRS~4) is detected in radio, suggests a correlation between the
outflow activity and the amount of circumstellar matter. The radio
properties of IRS~2 and IRS~4, and the quietness of IRS~6, conform
with the concept that the radio activity and the nature of the
emission change in the course of protostellar evolution, as discussed
e.g.\ by Gibb ({\cite{gibb99}) and Feigelson \& Montmerle
(\cite{feigelson99}).  The spectral index of radio emission from
CHX10a could not be determined because it lies outside the FOV at
3.5\,cm. As the star is a Class~III object associated with X-ray
emission, the 6\,cm emission is probably non-thermal.

The Class~0 candidate ChaMMS-1 (Reipurth et~al.\ \cite{reipurth96}) is
not detected in radio. Thus, since there is no clear evidence for jets
or outflow associated with this source, we suggest that it represents
a still earlier protostellar stage or a pre-stellar clump.  As a
Class~I protostar with optically thick ionized wind IRS~4 is the most
likely candidate for the central source of the molecular outflow
detected by Mattila et~al.\ (\cite{mattila89}).

Eight radio sources with no counterpart in other wavelengths
were detected in the region of Ced~110. Two of these additional
sources were detected at 3.5\,cm only and can be shocks associated with
the star forming region. The rest have negative spectral indices and
are probably radio galaxies.  The brighest of these is a compact
source with a spectral index of $-1.1$ and a weak linear polarization
($\sim 2\%$) at 6\,cm. Further radio observations of this source 
seem warranted.

\begin{acknowledgements}
We wish to thank the ATCA staff for their help during our
observations. We are grateful to Professor Kalevi Mattila for helpful
comments on the manuscript.  The work of K.L., S.K.\ and J.H.\ has
been supported by the Finnish Academy through grants Nos.\ 173~727 and
174854, which is gratefully acknowledged. The work of S.K.\ has been
supported by the V\"ais\"al\"a Foundation of the Finnish Academy of
Science and Letters.  This research has made use of the SIMBAD
database, operated at CDS, Strasbourg, France.
\end{acknowledgements}

\end{document}